\documentclass[prb,preprint,aps]{revtex4-1}

\usepackage{amsmath}
\usepackage{graphicx}

\newcommand{\grad}{\boldsymbol{\nabla}}
\newcommand{\Sv}{{\mathbf S}}

\begin{document}

\title{Poynting vector flow in a circular circuit}

\author{Basil S. Davis}
\email{bdavis2@tulane.edu}
\affiliation{Tulane University, Department of Physics, New Orleans,
Louisiana 70118}

\author{Lev Kaplan}
\email{lkaplan@tulane.edu}
\affiliation{Tulane University, Department of Physics, New Orleans,
Louisiana 70118}


\begin{abstract}
A circuit is considered in the shape of a ring, with a battery of negligible size and a wire of uniform resistance. A linear charge distribution along the wire maintains an electrostatic field and a steady current, which produces a constant magnetic field. Earlier studies of the Poynting vector and the rate of flow of energy considered only idealized geometries in which the Poynting vector was confined to the space within the circuit. But in more realistic cases the Poynting vector is nonzero outside as well as inside the circuit. An expression is obtained for the Poynting vector in terms of products of integrals, which are evaluated numerically to show the energy flow. Limiting expressions are obtained analytically. It is shown that the total power generated by the battery equals the energy flowing into the wire per unit time.
\end{abstract}

\maketitle

\section{Introduction}
\label{sec:intro}
Although the concept of the Poynting vector has been known for well over a century,\cite{Poynting} the actual flow of electromagnetic energy in even the simplest three-dimensional ohmic circuit is a subject of ongoing interest.\cite{Harbola} In this paper we calculate the electric and magnetic fields generated by a circular circuit carrying a steady current, and obtain the Poynting vector $ \Sv = \epsilon_0 c^2 \mathbf{E}\times \mathbf{B}$ representing the flow of electromagnetic energy in the space surrounding the circuit. We shall show that the energy flows through every point in space while traveling from the battery to the wire, and yet the net flow of energy into the resistor equals the energy generated by the battery.

\section{Two-dimensional circuits}
\label{sec:twodim}
\begin{figure}[h!]
\centering
\includegraphics[width=3in]{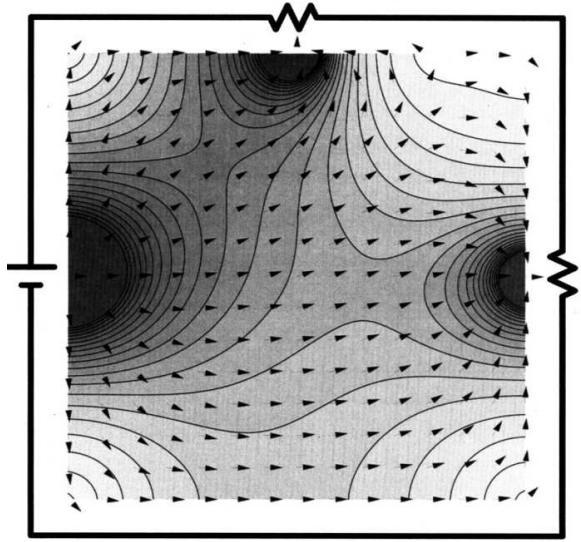}
\caption {Poynting vector for a two-dimensional circuit.\cite{Majcen}}
\label{fig:DavisKaplanFig01}
\end{figure}

The idea that electromagnetic energy flowing from a battery to a resistor in a simple circuit travels through the surrounding space rather than through the wires is counterintuitive.\cite{GaliliandGoihbarg} The calculation of the Poynting vector field also poses computational challenges when applied to a realistic three-dimensional circuit. Although realistic three-dimensional geometries have not been considered in detail, the case of a circuit consisting of an infinitely long shaft, which reduces the problem to two dimensions, has been treated with success.\cite{Majcen} It was shown that the magnetic field vanishes outside the circuit, and hence the Poynting vector vanishes as well. Therefore, for a two-dimensional geometry the energy flow is totally confined within the circuit. Figure~\ref{fig:DavisKaplanFig01} shows the two-dimensional section of the circuit studied in Ref.~\onlinecite{Majcen}. The battery, resistors, and wire are taken to be infinite in extent perpendicular to the plane of the figure. The arrows show the direction of the flow of electromagnetic energy, which is the Poynting vector field. The energy flows from the battery to the two resistors, the flow being entirely contained within the interior space of the rectangular shaft. However, the three-dimensional problem is qualitatively different from the two-dimensional one, because in the former $\Sv$ is no longer confined within the space defined by the circuit.\cite{Preyer}

\begin{figure}[h!]
\centering
\includegraphics[width=3in]{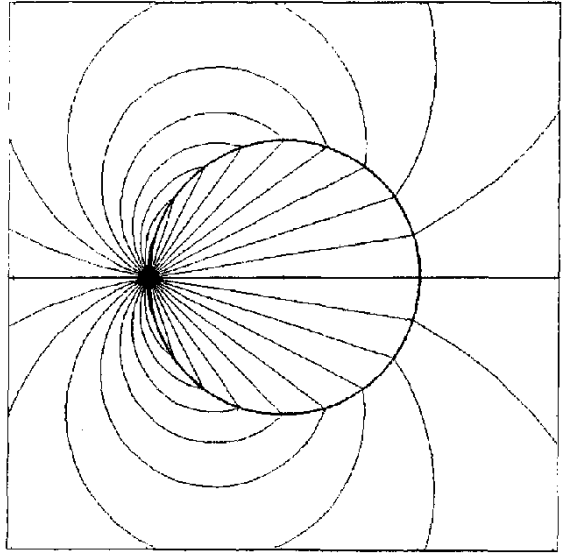}
\caption { Equipotentials for a two-dimensional circuit with a battery on the left and a uniform distributed resistance in the circular wire.\cite{Heald} The lines {inside the circuit} also represent the direction of the Poynting vector, which is directed away from the battery. The Poynting vector vanishes outside the circuit.}
\label{fig:DavisKaplanFig02}
\end{figure}

A related study examined the fields and energy flow in circuits where the two-dimensional cross section is a circle.\cite{Heald} Figure~\ref{fig:DavisKaplanFig02} shows the equipotential lines for such a circuit. These lines also represent the direction of $\Sv$, but only the lines in the interior of the circuit, because $\Sv$ vanishes on the outside. Finally, Ref.~\onlinecite{Jackson1} explored the distribution of charge in a coaxial circuit with a battery on the outer cylinder and a resistor on the inner cylinder and showed that $\Sv$ is nonzero only within the space between the two cylinders. Because of the peculiar shape of the circuit considered in Ref.~\onlinecite{Jackson1}, the results are not typical of a realistic circuit. Thus the idealized circuits studied in the literature share the feature that the Poynting vector is confined to the geometrical space of the circuit.

\section{A Three-dimensional circuit}
\label{sec:threedim}
\begin{figure}[h!]
\centering
\includegraphics[width=3in]{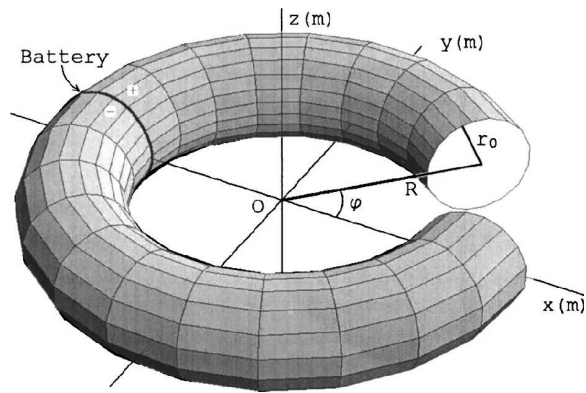}
\caption {Toroidal circuit.\cite{Hernandes} (Copyright (2003) by the American Physical Society.)}
\label{fig:DavisKaplanFig03}
\end{figure}

Hernandes and Assis\cite{Hernandes} have done a detailed treatment of a circuit in the form of a torus (see Fig.~\ref{fig:DavisKaplanFig03}). These authors calculated the electric fields inside and outside as well as the surface charge distribution on the conductor. The problem of finding the electric field due to a conductor carrying a steady current --- ``Merzbacher's puzzle''\cite{Merzbacher} --- has aroused much interest.\cite{Jackson1,AssisCisneros,griffith,jefimenkobook}

Although Ref.~\onlinecite{Hernandes} did not consider the magnetic field, their work is important for our purposes because an exact solution was provided for the electric field due to a three-dimensional ohmic circuit consisting of a battery driving a current through a resistor. Given the exact solution for a circuit in the shape of a toroid, Hernandes and Assis allowed the toroid to collapse to a ring-shaped wire whose thickness is small compared to the diameter of the ring. They thus obtained an expression for the surface charge density in a circular circuit made of a wire of small thickness. We shall calculate the surface charge density using a different method, and we will find that in the limit as the radius of the wire approaches zero, our result is in agreement with theirs.\cite{Hernandes}

In this paper we will consider a circuit in the shape of a ring with a battery of negligible size and a wire of uniform resistance per unit length (see Fig.~\ref{fig:DavisKaplanFig05}). We shall obtain expressions for the corresponding electric and magnetic fields, and plot $\Sv$ inside and outside the circuit. We will show that $\Sv$ of a simple circuit is not confined to the space within the circuit, and spans the space outside the circuit as well, even though the magnitude of $\Sv$ decreases as the sixth power of the distance from the circuit.

The exact value of $\Sv$ at an arbitrary point inside or outside the circuit can be obtained only by numerical integration. It will be shown analytically that the total amount of energy flowing from the battery into the wire equals the potential difference between the ends of the battery multiplied by the current flowing through the wire.

For a DC circuit consisting of a battery of constant emf and a thin wire of uniform resistance, it has been shown that charges develop all along the wire from one terminal of the battery to the other, and that the charge per unit length of the wire is a linear function of the distance from the battery.\cite{Jefimenko,jefimenkobook,Jackson1} The charge density calculation will allow us to obtain the electric field and hence the Poynting vector associated with the circuit.

\section{The surface charge density}
\label{sec:surfcharge}

\begin{figure}[h!]
\centering
\includegraphics[width=3in]{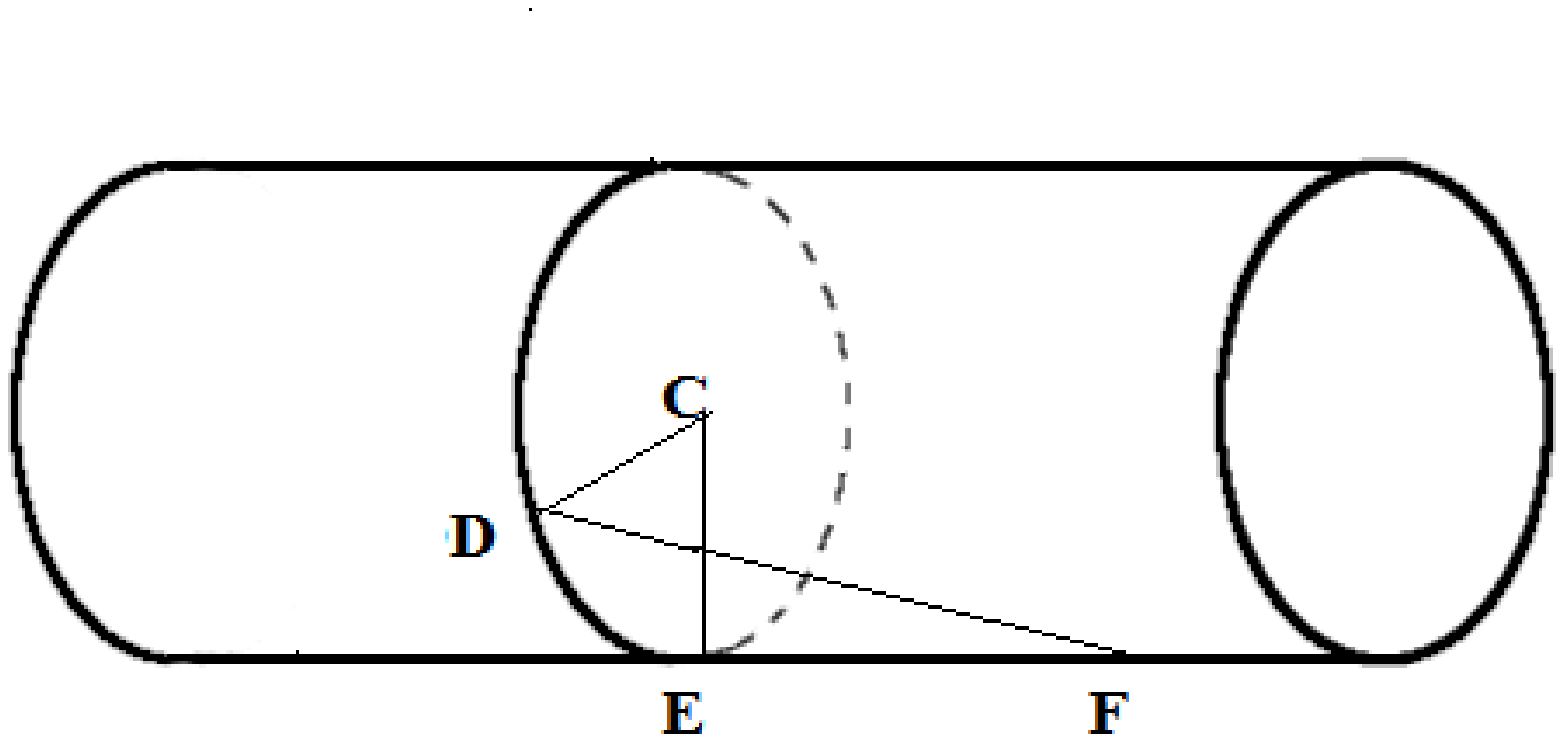}
\caption { Short segment of the wire. Point F is located at the linear coordinate $\xi_0$, and point E at $\xi$; $\mbox{CE} = \mbox{CD} = r_0$ is the radius. The arc ED subtends the angle $\psi$.}
\label{fig:DavisKaplanFig04}
\end{figure}

We first obtain an expression for the surface charge density $\sigma$ on the wire. To this end we shall assume that the radius of the wire $r_0$ is much smaller than the radius $R$ of the circuit. Given this assumption, we consider a portion of the circuit sufficiently far from the battery. Because $R\gg r_0$, the arc of the torus may be approximated by a straight cylinder, as shown in Fig.~\ref{fig:DavisKaplanFig04}. We take the length of the cylinder to be $2L$, and coordinate $\xi$ along the length of the cylinder to vary from $-L$ to $+L$.

We also assume that the surface charge density $\sigma$ on the wire depends only on $\xi$, and varies linearly with $\xi$ (see, for example, Ref.~\onlinecite{Jefimenko}). Hence
\begin{equation}
\sigma = a + b \xi \,,
\label{sigmaab}
\end{equation}
where $a$ and $b$ are suitable constants.

We consider the electrostatic potential at the point F in Fig.~\ref{fig:DavisKaplanFig04} located at $\xi=\xi_0$ on the surface of the cylinder, due to the total charge distribution on the cylinder.
We first determine the potential at F due to the charge on a ring of width $d\xi$ at point E.
Consider point D on the ring such that the arc ED subtends the angle $\psi$ at the center.
The potential at F due to the charge on an infinitesimal area element located at D is given by
\begin{equation}
dV(\xi_0) = { (a + b\xi)r_0\, d\psi \, d\xi \over 4\pi \epsilon_0 \sqrt{(\xi-\xi_0)^2 +\left(2r_0 \sin\dfrac{\psi}{2}\right)^2}} \,.
\end{equation}
The potential at F due to the entire cylindrical surface of length $2L$ is then
\begin{equation}
V(\xi_0) = {2\over{4\pi}\epsilon_0}\!\int_{0}^{\pi} d\psi\int_{-L}^{L}d\xi { (a + b\xi)r_0\over \sqrt{(\xi-\xi_0)^2 +\left(2r_0\sin\dfrac{\psi}{2}\right)^2}} \,,
\end{equation}
and the electric field parallel to the conductor at the surface of the cylinder is \begin{equation}E_\parallel(\xi_0) = -{{\partial V}\over{\partial \xi_0 }} \,.
\end{equation}
Because the charge density varies linearly with coordinate $\xi_0$, to obtain the constants $a$ and $b$ in Eq.~(\ref{sigmaab}) it suffices to determine the electric field at a point midway between the two ends of the cylinder, that is, at $\xi_0 = 0$.
A straightforward integration yields
\begin{equation}
E_\parallel ={ 2 br_0\over{4\pi \epsilon_0}} \int_{0}^{\pi} d \psi \left[ \frac{2\zeta }{\sqrt{1+\zeta^2 }}-\ln\left (\frac{\zeta+\sqrt{1+\zeta^2}}{-\zeta +\sqrt{1+\zeta^2}}\right)\right] \,,
\label{epar}
\end{equation}
where $\zeta=L/[2r_0 \sin(\psi/2)]$.

We now apply these results to a circuit in the form of a circle of radius $R$. The long straight conductor of length $2L$ becomes a curved wire of length $2\pi R$, so we will replace $L$ by $\pi R$ in the subsequent calculations. The curvature of the wire introduces a correction of order $1/\ln{(R/r_0)}$ into the expression we will obtain for the charge density needed to obtain a given potential difference $V$ between the two ends of the wire (Eq.~(\ref{sigmaxi}))~\cite{Hernandes}, and a correction of order $r_0/R$ into Eq.~(\ref{epar}) for the parallel electric field.
In particular, a variation in the charge density is necessary to guide the current in a circle rather than a straight line.
(In Sec.~\ref{sec:special} we will see that for the case of an infinitely thin circular wire,
the curvature correction to the linear charge density as a function of the external potential difference is $O(1/\ln{(R/h)})$, and the correction to the parallel electric field is
$O(h/R)$, where the potential and field are measured at a distance $h$ from the wire.)

In the following we will present our results for a thin wire, that is, keeping only the leading terms for $R \gg r_0$. 
Replacing $L$ with $\pi R$, we have $\zeta=\pi R/(2 r_0 \sin(\psi/2))$ in Eq.~(\ref{epar}).
Because $R \gg r_0$, it follows that $\zeta^2 \gg 1$.
We expand $1/\sqrt{1+\zeta^2 }$ in Eq.~(\ref{epar}) in powers of $1/\zeta^2$, and neglect $1/\zeta^4$ and higher order terms. We use the identity
\begin{equation}
\int_{0}^{\pi}\! d\psi \ln[\sin(\psi/2)]= -\pi \ln 2\,,
\end{equation}
and obtain the electric field parallel to the wire,
\begin{equation} E_\parallel(\xi) = - \frac{br_0}{\epsilon_0}\ln\left (\frac{2\pi R}{er_0}\right ) \,.
\end{equation}
Because $E_\parallel(\xi_0)$ is independent of $\xi_0$, the potential difference $V$ between the two terminals of the battery is $V=(2 \pi R)E_\parallel$, and hence
\begin{equation}
b = \frac{V \epsilon_0}{2\pi Rr_0\ln\left(\dfrac{2\pi R}{er_0}\right)} \,.
\end{equation}

In Eq.~(\ref{sigmaab}) the charge density varies linearly with $\xi$, and is positive close to the positive terminal of the battery, $\xi=+L=+\pi R$, and negative near the negative terminal, $\xi=-L=-\pi R$. By symmetry, the charge density must vanish half way between the terminals, where $\xi = 0$. Thus $a$ = 0, and
\begin{equation}
\sigma(\xi) = \frac{V \epsilon_0 \xi}{2\pi Rr_0\ln\left(\dfrac{2\pi R}{er_0}\right)} \,.
\label{sigmaxi}
\end{equation}

For a circular circuit in the $x$-$y$ plane with its center at the origin, $\xi$ can be replaced by the azimuthal variable $\phi$, where
\begin{equation} \xi = R(\phi -\pi)\,.
\end{equation}
We locate the battery of negligible length on the positive $x$ axis, with the negative terminal at $\phi = 0$ and the positive terminal at $\phi = 2\pi$. In these coordinates we have
\begin{equation}
\sigma(\phi) = V \epsilon_0 (\phi-\pi)/2\pi r_0\ln\left(\frac{2\pi R}{er_0}\right)\,,\end{equation}
which in the limit of a thin wire, $r_0/R\rightarrow 0,$ has the leading behavior
\begin{equation} \sigma(\phi) = \frac{V \epsilon_0 (\phi-\pi)}{2\pi r_0\ln(R/r_0)} \,.
\label{sigmaphi}
\end{equation}
This limiting form agrees with the thin wire limit of the result in Ref.~\onlinecite{Hernandes} for the surface charge density on a toroidal conductor carrying a steady current.

\section{Poynting vector in the near vicinity of the wire}
\label{sec:PVnearvicinity}
We next obtain limiting expressions for the Poynting vector very close to the wire, i.e., at distances small compared to the wire circumference $2 \pi R$. For such calculations the curvature of the circuit becomes negligible, and we may approximate the circuit by an infinitely long cylindrical conductor. We previously made such an approximation in determining the field parallel to the wire to obtain an expression for the surface charge density. We will also need expressions for the electric field normal to the conductor, that is, the radial electric field, and the magnetic field circulating the wire. By combining the magnetic field with the radial and longitudinal electric fields, we will obtain both components of the Poynting vector near the wire.

The expression for the radial electric field outside a charged cylindrical wire is given by\cite{Marcus}
\begin{equation}
E_\perp(r) = \frac{r_0\sigma}{\epsilon_0 r}\,, \label{Eperp}\end{equation}
where $r>r_0$ is the distance from the axis of the cylinder.
Similarly, the magnetic field outside and close to a cylinder carrying a steady current $I$ is given by
\begin{equation}
B =\frac{I}{2\pi\epsilon_0 c^2 r}\,.
\label{bfield}
\end{equation}
This magnetic flux circulates azimuthally around the cylinder, and hence the magnetic field is normal to both the radial and the longitudinal electric fields.
Thus, the Poynting vector parallel to the conductor at a distance $r$ from the axis is \begin{equation}
S_\parallel= \epsilon_0 c^2 E_\perp B = \frac{r_0\sigma I}{2\pi\epsilon_0 r^2}=\frac{VI}{4\pi r^2} \frac{\phi-\pi}{\ln\left(\dfrac{2\pi R}{e r_0}\right)} \,,\label{Sparl}
\end{equation}
and the total energy flowing per second through an annulus of inner radius $r_0$ and outer radius $r$, along the direction of the current, is given by
\begin{subequations}
\begin{align}
P_\parallel(r_0,r)&=\frac{r_0 \sigma I}{\epsilon_0}\ln (r/r_0) \\
& = \frac{VI}{2} \frac{\ln(r/r_0)}{\ln(R/r_0)}\left( \frac{\phi}{\pi}-1\right) \,, \label{second}
\end{align}
\end{subequations}
where in Eq.~\eqref{second} we have used Eq.~(\ref{sigmaphi}) for the surface charge density.

\begin{figure}[h!]
\centering
\includegraphics[width=3in]{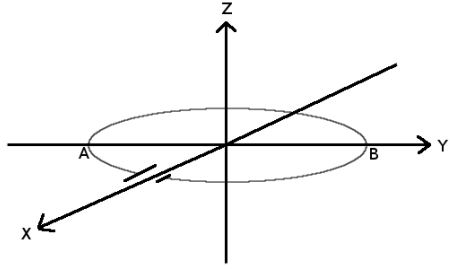}
\caption {The circuit intersects the $y$-$z$ plane at points A and B.}
\label{fig:DavisKaplanFig05}
\end{figure}

Consider, for example, points A and B at which the circuit intersects the $y$ axis (see Fig.~\ref{fig:DavisKaplanFig05}). At A the azimuthal angle is $\phi = 3\pi/2$, and thus the power flowing through a small annular region around the wire at A is given by
\begin{equation}
P_{\parallel,A}=+\frac{VI}{4} \frac{\ln(r/r_0)}{\ln(R/r_0)}\,.
\label{17}
\end{equation}
At A the current flows into the $y$-$z$ plane, and $P_{\parallel, A}$ is positive, indicating energy flow along the direction of the current, that is, away from the battery. At point B, the corresponding angle is $\phi = \pi/2$, and the rate of energy flow is given by
\begin{equation}
P_{\parallel, B}=-\frac{VI}{4} \frac{\ln(r/r_0)}{\ln(R/r_0)}\,.
\label{18}
\end{equation}
The magnitude of the Poynting vector at points A and B is the same, due to symmetry.
The negative sign at point B indicates the energy is being transported { opposite} to the direction of current flow, that is, into the $y$-$z$ plane and away from the battery, as at A.

We now let the wire radius $r_0\rightarrow 0$ keeping $r$ and $R$ finite. Equations~\eqref{17} and \eqref{18} for $P_{\parallel,A}$ and $P_{\parallel,B}$ lead to $\pm VI/4$, and
thus the energy flowing across the $y$-$z$ plane through a pair of circles of radius $ r\gg r_0$ centered at A and B is
\begin{equation}
P_{\rm yz \; plane}=\frac{VI}{2}\,. \label{20}
\end{equation}
The result in Eq.~\eqref{20} is equal to half the energy produced per second by the battery, because the $y$-$z$ plane cuts the circuit in half. Now $r$ can take any arbitrarily small nonzero value, as long as $r_0\rightarrow 0$. So for a wire of negligible thickness, almost all of the energy transport described by the Poynting vector occurs inside an arbitrarily thin cylinder surrounding the wire.

To investigate the energy transport very close to the surface of the wire, we consider an annular region $r_0 \le r \le r_0 + h,$ where $0<h\ll r_0 \ll R$.
The power transported through this annular region becomes
\begin{equation}
P_\parallel(r,r_0)=\frac{VI}{4}\frac{h/r}{\ln(R/r_0)}\left( \frac{\phi}{\pi}-1\right)\,. \label{21}
\end{equation}
Equation~\eqref{21} shows that the amount of energy flowing through an annular region of thickness $h$ near the surface of the wire increases linearly with $h$. Most of the total power in the circuit is transported at distances $h \sim r_0 \ln(R/r_0)$ from the wire of radius $r_0$, that is, at distances moderately larger than the wire radius. The Poynting vector drops off as the inverse square of the distance from the axis of the wire, $r=r_0+h$, as shown in Eq.~(\ref{Sparl}).

We now turn to the other component of the Poynting vector near the wire. If we combine Eq.~(\ref{epar}) for the parallel electric field and Eq.~(\ref{bfield}) for the circulating magnetic field in the vicinity of the wire, we find the component of the Poynting vector perpendicular to the wire to be
\begin{equation}
S_\perp = \frac{VI}{4 \pi^2 R r}\,.\label{Sperp}
\end{equation}
Therefore the power per unit length flowing into the wire is
\begin{equation}
P_\perp= \frac{VI}{2 \pi R}\,, \label{above}
\end{equation}
which is the same at all points along the wire.

It can be readily shown that Eq.~\eqref{above} becomes identical with Eq.~(4) of Ref.~\onlinecite{Harbola} for a point very close to the surface of a straight wire. If we integrate over the entire circuit, the total energy flowing per second into the wire is $VI$, as expected.

\section{Electric field due to a circular circuit}
\label{sec:efieldcirc}

\begin{figure}[h!]
\centering
\includegraphics[width=3in]{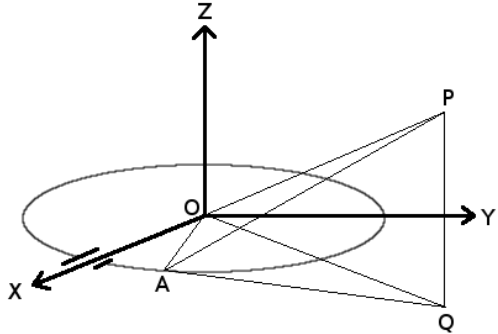}
\caption {Circuit with uniform resistance.}
\label{fig:DavisKaplanFig06}
\end{figure}

In Sec.~\ref{sec:PVnearvicinity} we analyzed the behavior of the Poynting vector and energy flow in the vicinity of the wire. We now address the Poynting vector in general, still assuming that the radius $R$ of the circuit is very large compared to the wire radius $r_0$. The following equation relates two different expressions for the charge on an element of the wire:
\begin{equation} 2\pi r_0 \sigma d\xi= \frac{V\epsilon_0 R}{\ln(R/r_0)}(\phi-\pi) d\phi \,,\end{equation}
where we have used Eq.~(\ref{sigmaphi}) for the surface charge density $\sigma$ on the wire. It is convenient to define $\lambda \equiv V\epsilon_0 R/ \ln(R/r_0)$. The electrostatic potential at an arbitrary point $ P=(r, \theta, \phi )$ in spherical coordinates (see Fig.~\ref{fig:DavisKaplanFig06}) is given by
\begin{equation}
\Phi(r, \theta, \phi) =\frac{\lambda}{4\pi \epsilon_0}\int_{0}^{2\pi}\!\frac{(\phi'-\pi)d\phi' }{\sqrt{R^2 +r^2 -2Rr \sin\theta \cos(\phi -\phi')}}\,,
\label{ScalarP}\end{equation}
and in Cartesian coordinates
\begin{equation}
\Phi(x, y, z) =\frac{\lambda}{4\pi \epsilon_0}\!\int_{0}^{2\pi}\!\frac{(\phi'-\pi)d\phi' }{\sqrt{R^2 +x^2+y^2+z^2 -2Rx\cos\phi' - 2Ry\sin\phi'}}\,.
\end{equation}
Hence
\begin{equation}
\textbf{E}(x,y,z) =\frac{\lambda}{4\pi \epsilon_0}\!\int_{0}^{2\pi}\!d\phi' \frac{(\phi'-\pi)[(x-R\cos\phi')\,\hat {\bf x}+ (y-R \sin\phi')\, \hat {\bf y}+z\, \hat {\bf z}]}{(R^2 +x^2+y^2+z^2 -2Rx\cos\phi' - 2Ry\sin\phi')^{3/2}} \label{E_vec} \,.
\end{equation}
The magnetic field at any point may be readily obtained as $\textbf{B}= \grad \times \textbf{A}$ from the vector potential\cite{Jackson3}
\begin{equation}
A_\phi (r, \theta,\phi) = \frac{\mu_0IR}{4\pi}\!\int_{0}^{2\pi}\!\frac{\cos \phi' d\phi'}{(R^2 +r^2 -2Rr \sin \theta \cos \phi')^{1/2} }\,.\label{Vectorpotential}
\end{equation}

\section{Special cases}
\label{sec:special}

To demonstrate that the curvature of the wire contributes a small correction to the Poynting vector for a large circuit of radius $R$, we consider an infinitely thin wire. Let us take a point on the $x$ axis at a small distance $h$ from the wire, that is, at a distance $R+h$ from the origin along the negative $x$ axis. The coordinates of this point are $(-R-h, 0, 0)$. We evaluate Eq.~(\ref{E_vec}) and find that that the electric field at $(-R-h, 0, 0)$ is parallel to the $y$ axis and is given by
\begin{subequations}
\begin{align}
E_y(-R-h, 0,0) &= \frac{\lambda}{4\pi \epsilon_0 R^2}\left[2 \ln \left(\frac{R}{h}\right) -\pi + 3 \ln 4 -\frac{h}{R} + 3 \frac{h}{R} \ln\left(\frac{h}{R}\right) + O \left(\frac{h^2}{R^2}\right)\right] \label{E_y} \\
&= \frac{\lambda}{2\pi \epsilon_0 R^2}\ln\left(\frac{8R}{e^{\pi/2}h}\right)\left[1+O\left(\frac{h}{R}\right)\right]\,.
\label{thincircle}
\end {align}
\end{subequations}
We can perform the same calculation for a point at a distance $h$ from the center of a straight infinitely thin wire of length $2 \pi R$. The result is
\begin{subequations}
\begin{align}
E_\parallel &= \frac{\lambda'}{4\pi \epsilon_0 R^2}\left [2 \ln \left(\frac{R}{h}\right) +2 \ln (2\pi) - 2 + O\left(\frac{h^2}{R^2}\right)\right] \\
&= \frac{\lambda'}{2\pi \epsilon_0 R^2}\ln\left(\frac{2\pi R}{e h}\right)\left[1+O\left(\frac{h^2}{R^2\ln(R/h)}\right)\right]\,.
\label{thinstraight}
\end{align}
\end{subequations}
If we compare Eqs.~(\ref{thincircle}) and (\ref{thinstraight}), we must equate the leading terms because we are assuming the thin circular wire and the thin straight wire are connected to batteries of the same voltage. We see that the curvature correction is given by the $O(h/R)$ term, which is present for the infinitely thin circular wire and absent for the infinitely thin straight wire. This curvature correction is analogous to the $O(r_0/R)$ curvature correction for a toroidal wire of radius $r_0$. We also note that $\lambda=\left[1+O\left(1/\ln R/h\right)\right]\lambda'$, that is, the charge density required to produce a voltage drop $V$ along a circular wire of length $2 \pi R$ is the same as the charge density required to produce the same voltage drop along a straight wire of the same length as $R \to \infty$.

We next consider what happens at the center of the circle, that is, at the origin $(0, 0, 0)$. Here we have \begin{equation}\textbf{E} = \frac{\lambda}{2\epsilon_0 R^2}\, \hat {\bf y} =\frac{V}{2R\ln(R/r_0)}\,\hat {\bf y}\,,
\end{equation}
and the magnetic field is given by
\begin{equation} \textbf{B} = -\frac{\mu_0 I}{2R}\hat {\bf z}\,.
\end{equation}
Thus the Poynting vector at the center of the circuit is directed along a diameter leading away from the battery and has the magnitude
\begin{equation}
S=\frac{VI}{4R^2 \ln(R/r_0)}\,. \label{thisexp}
\end{equation}
As $r_0\rightarrow 0,$ the right-hand side of Eq.~\eqref{thisexp} approaches 0. This behavior is in agreement with our earlier result that for a very narrow wire, the energy flow stays very close to the wire, rapidly dropping to zero far from the wire.

We now consider the behavior of the Poynting vector far from the origin. At distances $r \gg R$, the electric field
falls off as $1/r^3$ and takes the asymptotic form
\begin{equation}
\textbf{E}(\textbf{r})= \frac{3\hat {\bf r}(\textbf{p}\cdot\hat {\bf r})-\textbf{p}}{4\pi\epsilon_0r^3}\,,\end{equation}
where $\textbf{p}$ is the electric dipole moment of the circuit. From Eq.~(\ref{E_vec}) we obtain
\begin{equation}
\textbf{p} = -\frac{2\pi V\epsilon_0R^2}{\ln(R/r_0)} \,\hat {\bf y}\,,
\end{equation}
where the negative sign indicates that the moment is directed along the negative $y$ axis. Thus the electric field may be expressed as
\begin{equation}
\textbf{E}(\textbf{r})= \frac{ VR^2}{\ln(R/r_0)} \frac{r \hat {\bf y}-3y\hat {\bf r}}{2r^4}\,. \label{elarger}
\end{equation}
Equation~\eqref{elarger} can also be obtained directly from the scalar potential in Eq.~(\ref{ScalarP}). The same expression is obtained by using the definition of a dipole moment with the charge density given by Eq.~(\ref{sigmaxi}). The magnetic field for $r \gg R_0$ similarly drops off as the inverse cube of the distance, and is given by
\begin{equation} \textbf{B}(\textbf{r}) = \frac{\mu_0R^2I(r \hat {\bf z}-3z \hat {\bf r})}{4r^4} \,. \label{blarger} \end{equation}
We combine Eqs.~(\ref{elarger}) and (\ref{blarger}) and find
\begin{equation}
\Sv =\epsilon_0 c^2 \textbf{E} \times \textbf{B}= \frac{VI R^4}{8r^6 \ln(R/r_0)}[(1-3\frac{y^2 }{r^2 } - 3\frac{z^2 }{r^2 })\, \hat {\bf x}+ (\frac{3xy}{r^2 })\, \hat {\bf y}+ (\frac{3xz}{r^2 })\,\hat {\bf z}] \,,
\end{equation}
or equivalently,
\begin{equation}
\Sv = \frac{VI R^4}{8r^6 \ln(R/r_0)}[(1- 3 \sin^2\theta \sin^2 \phi - 3 \cos^2\theta)\,\hat {\bf x}+ (3 \sin^2\theta \sin\phi \cos\phi)\,\hat {\bf y}+ (3 \sin\theta \cos\theta \cos\phi)\,\hat {\bf z}] \,. \label{largedist}
\end{equation}
We have obtained the important result that at sufficiently large distances from the circuit, the Poynting vector drops off as the inverse sixth power of the distance. We also observe that at any point on the three coordinate axes $\Sv$ is directed along or parallel to the $x$ axis.

We now consider $\Sv$ in the $x$-$y$ plane containing the circular wire,
where the Poynting vector may be written as
\begin{equation}
\Sv=\epsilon_0 c^2( E_y B_z \hat {\bf x}- E_x B_z \hat {\bf y}) \,,
\end{equation}
where $\textbf{E}$ in the $x$-$y$ plane is given by Eq.~(\ref{E_vec}) with $z=0$,
\begin{equation}
\textbf{E}(x,y) =\frac{VR}{4\pi \ln(R/r_0)} \int_{0}^{2\pi}d\phi' \frac{(\phi'-\pi)[(x-R\cos\phi')\,\hat {\bf x}+ (y-R \sin\phi')\,\hat {\bf y}]}{[R^2 +x^2+y^2 -2R(x\cos\phi' +y\sin\phi')]^{3/2}} \label{E_x2}
\,,\end{equation}
and $\textbf{B}$ by the curl of Eq.~(\ref{Vectorpotential}),
\begin{equation}
\textbf{B}(x,y)= -\frac{A_\phi + r\partial A_\phi/\partial r}{r}\hat {\bf z} = -\frac{\mu_0 I}{4 \pi r} \hat {\bf z}
\int_{0}^{2\pi}d\phi'\frac{\cos \phi' (1 -\dfrac{r}{R}\cos\phi')}{(1+ \dfrac{r^2}{R^2}-2\dfrac{r}{R}\cos\phi') ^{3/2} }\,.
\end{equation}

\begin{figure}[h!]
\centering
\includegraphics[width=3in]{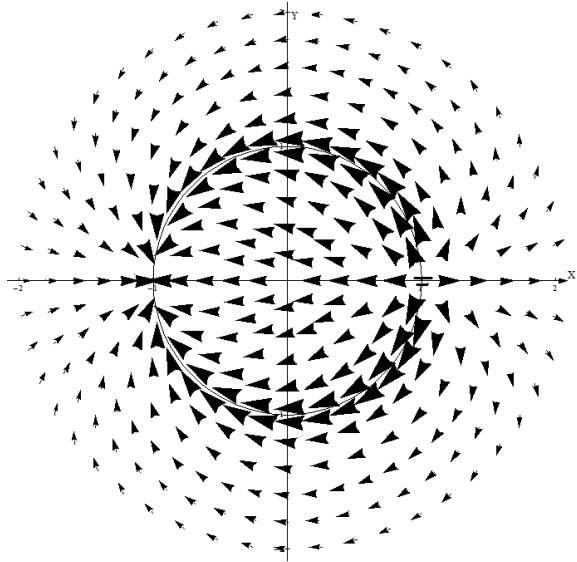}
\caption {Logarithmic plot of the Poynting vector in the plane of the circuit.}
\label{fig:DavisKaplanFig07}
\end{figure}

Figure~\ref{fig:DavisKaplanFig07} shows the Poynting vector for a circular circuit with uniform resistance.
The battery is on the $x$ axis.
We see from Eq.~(\ref{Sparl}) that the component of the Poynting vector parallel to the wire at a distance $r$ from the wire axis
is proportional to $1/r^2$. Near the surface of the wire, $r \sim r_0$, the parallel component $S_\parallel$ given by Eq.~(\ref{Sparl}) will be much greater than
the perpendicular component $S_\perp$ given by Eq.~(\ref{Sperp}), as long as the wire is thin ($r_0 \ll R$). Thus, almost everywhere the arrows in Fig.~\ref{fig:DavisKaplanFig07} are parallel to the circuit. Equation~(\ref{Sparl}) shows that $S_\parallel$ is also proportional to the charge density $\sigma$, which varies as $\phi - \pi$. Thus, in a region directly opposite to the battery, $\phi \approx \pi$, the parallel component $S_\parallel$ is small compared to the perpendicular component $S_\perp$, which has the same value at all points along the circuit, cf. Eq.~(\ref{Sperp}). It follows that close to the half way point of the circuit, the arrows are perpendicular to the wire, as required by the symmetry of the circuit.

\begin{figure}[h!]
\centering
\includegraphics[width=3in]{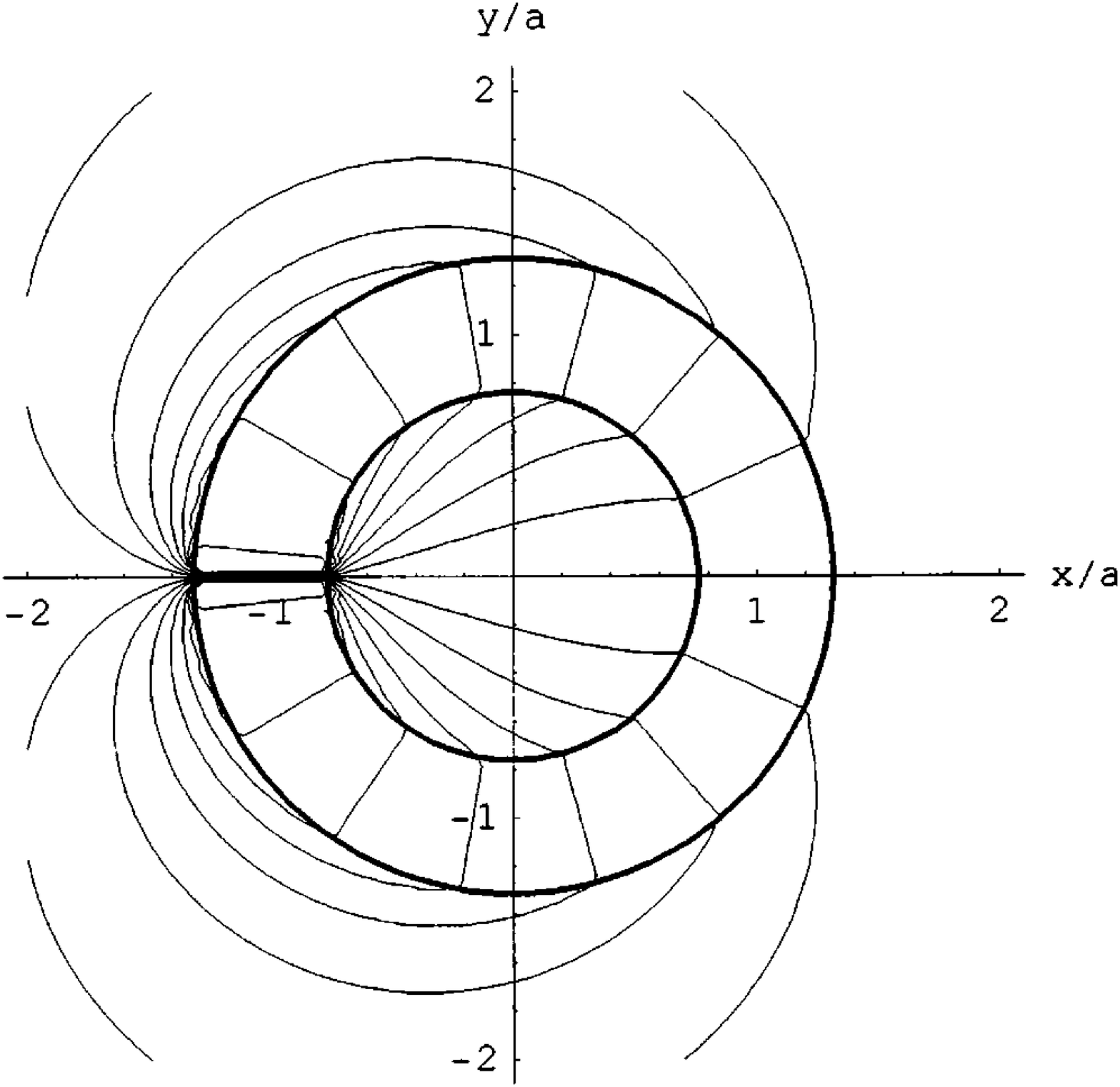}
\caption {Equipotential lines for a toroidal circuit.\cite{Hernandes} (Copyright (2003) by the American Physical Society.) The battery is located on the left, and the equipotential lines coincide with the direction of the Poynting vector field, which points outward from the battery. Within the body of the toroid the energy flows inward from the outer surface and outward from the inner surface, decreasing to zero somewhere inside.}
\label{fig:DavisKaplanFig08}
\end{figure}

Figure~\ref{fig:DavisKaplanFig08} shows the equipotential lines in the toroidal circuit investigated in Ref.~\onlinecite{Hernandes}. In Fig.~\ref{fig:DavisKaplanFig08} the torus represents a circuit with a battery on the left, and uniform resistance per unit length.
The equipotential lines coincide with the direction of the Poynting vector at every point in the $x$-$y$ plane, because both are normal to the electric field in the $x$-$y$ plane passing through that point. Thus Fig.~\ref{fig:DavisKaplanFig08} shows the direction but not the magnitude of the Poynting vector inside and outside the circuit. In the empty space surrounding the (solid) toroid, the Poynting vector is directed from the battery to points on the surface of the toroid. Within the body of the toroidal wire, assuming that the current is nonzero in the interior, the Poynting vector points from the surface of the wire to the interior of the wire. Specifically, the Poynting vector points inward from the outer surface at radius $R+r_0$ from the center of the circuit and outward from the inner surface at radius $R-r_0$, monotonically decreasing as it moves into the interior and coming to zero inside the wire. The pattern of electromagnetic energy flow for a toroidal circuit may be compared with the results we have obtained for a thin circular wire (Fig.~\ref{fig:DavisKaplanFig07}).

\section{Global integral}
\label{global}

We now calculate the total electromagnetic energy flowing from the battery into the diametrically opposite half of the circuit through the $y$-$z$ plane (see Fig.~\ref{fig:DavisKaplanFig05}).

For steady fields and currents we have $ \textbf{E} = -\grad \Phi$, so that the total power crossing the $y$-$z$ plane in the $-x$ direction is
\begin{subequations}
\begin{align} P &= \epsilon_0 c^2\!\int_{-\infty}^{\infty}\int_{-\infty}^{\infty} (\grad \Phi \times \textbf{B}) \cdot \hat {\bf x} \,dy\,dz \\
&= \epsilon_0 c^2 \int_{-\infty}^{\infty}\int_{-\infty}^{\infty} [\grad \times (\Phi \ \textbf{B})-\Phi\ (\grad \times \textbf{B})] \cdot \hat {\bf x} \, dy\, dz \,. \label{totpower}
\end{align}
\end{subequations}
The first term in the integrand in Eq.~(\ref{totpower}) equals
\begin{equation}
\epsilon_0 c^2\!\int_{-\infty}^{\infty}\int_{-\infty}^{\infty} \left[ \frac{ \partial (\Phi B_z) }{\partial y} - \frac{ \partial (\Phi B_y ) }{\partial z}\right] dy\,dz \,, \end{equation}
which reduces to zero because both the scalar potential and the magnetic field vanish at $\pm \infty$. From Maxwell's equations we have $ \epsilon_0 c^2 \grad \times \textbf{B} = \textbf{J}$ for steady fields and currents,
and thus the second term in Eq.~(\ref{totpower}) yields \begin{equation}
P = - \!\int_{-\infty}^{\infty} \int_{-\infty}^{\infty} \Phi\ \textbf{J}\cdot\hat {\bf x} \, dy\, dz \,. \end{equation}
The current density $\textbf{J}$ vanishes in the $y$-$z$ plane except in the two regions where the wire intersects the plane. These regions are centered at points A and B in Fig.~\ref{fig:DavisKaplanFig05}, and we have $ \textbf{J} \cdot \hat {\bf x}< 0 $ at A and $ \textbf{J}\cdot \hat {\bf x}> 0 $ at B.
Therefore
\begin{equation}
P = (\Phi_A - \Phi_B) I \,,
\end{equation}
where $\Phi_A $ is the electrostatic potential at A and $\Phi_B $ the potential at B. Note that in this analysis no assumptions were made regarding the specific shape of the circuit.

Now consider a closed surface of arbitrary shape that encloses the battery. Let C and D be the points at which the wires intersect the surface, with C being the high potential point and D the low potential point. The total energy flow per unit time through this surface is
\begin{equation}
P = \epsilon_0 c^2 \!\oint (\textbf{E}\times \textbf{B}) \cdot \textbf{n} \,dS \,,
\end{equation}
where the closed integral is over the surface and $\textbf{n}$ is the outward unit vector. Because $ \textbf{E} = - \grad \Phi $ for time-independent fields,
\begin{equation}
P = -\epsilon_0 c^2 \!\oint (\grad \times (\Phi \ \textbf{B})) \cdot \textbf{n} \,dS + \epsilon_0 c^2 \!\oint \Phi \ (\grad \times \textbf{B}) \cdot \textbf{n} \,dS \,.
\end{equation}
The first term vanishes by Gauss' divergence theorem. We apply Maxwell's equation for steady fields to the second term and obtain
\begin{equation}
P =\!\oint \Phi \textbf{J} \cdot \textbf{n} \,dS= (\Phi_C - \Phi_D) I.
\end{equation}
If the resistance is confined to a small region of the circuit (a ``lumped'' resistor), we can do a similar calculation with the closed surface enclosing the resistor, with the wire intersecting the surface at E and F. We then obtain the rate of energy flow into the resistor as
\begin{equation}
P = (\Phi_E - \Phi_F) I \,.
\end{equation}
We obtain the counterintuitive result that regardless of the shape of the circuit, the energy flows from the battery and enters the resistor after traveling through every point in space. The total rate of energy flow from the battery into the resistor is always $VI$, where $V$ is the potential difference across the ends of the resistor and $I$ the current flowing through the circuit.

\section{Concluding remarks}
\label{conclusion}

We have shown that the Poynting vector is not confined to the interior of the circuit, but flows through all space from the battery to the resistor. Part of the electromagnetic energy takes the shortest route, which is typically shorter than the distance along the wires. A small part of the energy follows very long paths from the battery to the wire. Maxwell's equations suggest that in an ordinary device such as a flashlight some of the energy makes a very long space odyssey from the battery to the bulb, exploring every cubic nanometer of space in the process. Consider an annular region in the $y$-$z$ plane (see Fig.~\ref{fig:DavisKaplanFig05}) with large inner radius $r \gg R$ and outer radius $r+dr$. From Eq.~(\ref{largedist}) we see that the amount of energy flowing per unit time across this annular area is proportional to $2\pi r \, dr/r^6 $,and thus the amount of energy flowing across the $y$-$z$ plane outside a large circle of radius $\rho$ centered on the origin is proportional to
\begin{equation}
\int_{\rho}^{\infty} \frac{r\, dr}{r^6 } \propto \frac{1}{\rho^4 }\,.
\end{equation}
An exact calculation may be easily performed using Eq.~(\ref{largedist}). We integrate the $x$ component of the Poynting vector over the entire $y$-$z$ plane outside a large radius $\rho$ and obtain
\begin{equation}
VI\frac{\pi}{8 \ln(\dfrac{R}{r_0})}\dfrac{R^4}{\rho^4}
\end{equation}
for the energy crossing the $y$-$z$ plane per unit time outside a large circle of radius $\rho$. For a circular wire with $r_0 = 1$\,mm and $R= 1$\,m, the energy per unit time crossing the $y$-$z$ plane outside a circle of radius $\rho = 10$\,m is $5.68\times 10^{-6} (VI/2)$, where $VI/2$ is the total amount of energy that crosses the entire $y$-$z$ plane per unit time.

\begin{acknowledgments}
This work was supported in part by the NSF under grant PHY-0545390.
\end{acknowledgments}



\begin{thebibliography}{99}
\bibitem{Poynting} J. H. Poynting, ``On the transfer of energy in an electromagnetic field,'' Phil. Trans. {\bf 175}, 277 (1884).

\bibitem{Harbola} M. K. Harbola,``Energy flow from a battery to other circuit elements: Role of surface charges,'' Am. J. Phys. \textbf{78}, 1203--1206 (2010). Harbola examined the Poynting vector flow due to a rectangular circuit, and suggested that similar studies be done for other geometries.

\bibitem{GaliliandGoihbarg} I. Galili and E. Goihbarg, ``Energy transfer in electric circuits: A qualitative account,'' Am. J. Phys. \textbf{73}, 141--144 (2005).

\bibitem{Majcen} S. Majcen, R. K. Haaland, and S. C. Dudley, ``The Poynting vector and power in a simple circuit,'' Am. J. Phys. {\bf 68}, 857--859 (2000).

\bibitem{Preyer} N. W. Preyer, ``Surface charges and fields of simple circuits,'' Am. J. Phys. \textbf{68}, 1002--1006 (2000). Preyer has plotted the Poynting vector flow for a rectangular circuit.

\bibitem{Heald} M. Heald, ``Electric fields and charges in elementary circuits,'' Am. J. Phys. \textbf{52}, 522--526 (1984).

\bibitem{Jackson1} J. D. Jackson, ``Surface charges on circuit wires and resistors play three roles,'' Am. J. Phys. \textbf{64}, 855--870 (1996).

\bibitem{Hernandes} J. A. Hernandes and A. K. T. Assis, ``Electric potential for a resistive toroidal conductor carrying a steady azimuthal current,'' Phys. Rev. E \textbf{68}, 046611-1--11 (2003). 

\bibitem{Merzbacher} E. Merzbacher, ``A puzzle from Professor Eugen Merzbacher,'' Am. J. Phys. \textbf{48}, 178 (1980).

\bibitem{AssisCisneros} A. K. T. Assis and J. I. Cisneros, ``The problem of surface charges and fields in coaxial cables and its importance for relativistic physics,''  in \textit{Open Questions in Relativistic Physics}, edited by F. Selleri (Apeiron, Montreal, 1998), pp. 177--185.

\bibitem{griffith} D. J. Griffith, \textit{Introduction to Electrodynamics}, 3rd ed. (Prentice Hall, Englewood Cliffs, NJ, 1999), pp. 336--337.

\bibitem{jefimenkobook} O. D. Jefimenko, \textit{Electricity and Magnetism}, 2nd ed. (Electret Scientific Company, Star City, WV, 1989).

\bibitem{Jefimenko} O. Jefimenko,``Demonstration of the electric fields of current-carrying conductors,'' Am. J. Phys. \textbf{30}, 18--21 (1962).

\bibitem{Marcus} A. Marcus, ``The electric field associated with a steady current in a long cylindrical conductor,'' Am. J. Phys. \textbf{9}, 225--226 (1941).

\bibitem{Jackson3} J. D. Jackson, \textit{Classical Electrodynamics}, 3rd ed. (John Wiley and Sons, Hoboken, NJ, 1999), Eq.~(5.36).

\end{thebibliography}
\end{document}